\documentclass[amsfonts,showpacs,tightenlines,aps,12pt,floatfix]{revtex4}
\usepackage{bm}
\usepackage{epsfig}
\begin{document}


\title{Isospin Breaking in the Relation Between 
the $\tau^-\rightarrow\nu_\tau\pi^-\pi^0$ and $e^+e^-\rightarrow \pi^+\pi^-$ 
Versions of $\vert F_\pi (s)\vert^2$ and Implications for $(g-2)_\mu$}
\author{Kim Maltman}
\email[]{kmaltman@yorku.ca}
\affiliation{Department of Mathematics and Statistics, York University, 
4700 Keele St., Toronto, ON CANADA M3J 1P3}
\altaffiliation{CSSM, Univ. of Adelaide, Adelaide, SA 5005 AUSTRALIA}
\author{Carl E. Wolfe}
\email[]{wolfe@yorku.ca}
\affiliation{Department of Physics and Astronomy, York University, 
4700 Keele St., Toronto, ON CANADA M3J 1P3}
\date{\today}

\begin{abstract}
We investigate two points related to existing treatments
of isospin-breaking corrections to the CVC relation 
between $\sigma [e^+e^-\rightarrow \pi^+\pi^-]$ 
and $d\Gamma [\tau^-\rightarrow\nu_\tau\pi^-\pi^0]/ds$. Implications
for the value of the hadronic contribution to $a_\mu =(g-2)_\mu /2$
based on those analyses incorporating hadronic $\tau$ decay data
are also considered.
We conclude that the uncertainty on the isospin-breaking
correction which must be applied to the $\tau$ decay data should be
significantly increased, and that the central value of the
$\rho$-$\omega$ ``mixing'' contribution to this correction
may be significantly smaller than indicated by the present
standard determination. Such a shift would contribute to
reducing the discrepancy between the 
$\tau$- and electroproduction-based determinations of the
leading order hadronic contribution to $a_\mu$.
\end{abstract}


\maketitle

\section{\label{intro}Introduction}
It is well known that, after the large, purely leptonic contribution,
the largest of the remaining Standard Model contributions 
to the anomalous magnetic moment of the muon, $a_\mu \equiv (g-2)_\mu /2$,
is that due to the leading order hadronic vacuum 
polarization, $\left[ a_\mu\right]_{had}^{LO}$. 
This contribution may be evaluated, in terms of experimental 
$e^+e^-\rightarrow hadrons$ cross-section
data, using the dispersion integral representation~\cite{gdr69}
\begin{equation}
\left[ a_\mu\right]_{had}^{LO}\, =\, {\frac{\alpha^2_{EM}(0)}{3\pi^2}}\,
\int_{4m_\pi^2}^\infty ds {\frac{K(s)}{s}}\, R(s)\ ,
\label{gm2weight}\end{equation}
where the form of $K(s)$ is well-known~\cite{gdr69} and $R(s)$ is the ratio of
the ``bare'' $e^+e^-\rightarrow hadrons$ cross-section to
that for $e^+e^-\rightarrow\mu^+\mu^-$~\cite{vpfootnote}.
Since the isovector part
of the electromagnetic (EM) spectral function is related by
CVC to the charged current isovector vector spectral function,
which can be obtained from the invariant mass distribution of
states with zero net strangeness in the decay
in $\tau^-\rightarrow\nu_\tau + hadrons$, 
the high precision hadronic $\tau$ decay data of 
Refs.~\cite{aleph97,opal99,cleo00} can,
in principle, be used to improve the determination of 
$\left[ a_\mu\right]_{had}^{LO}$~\cite{adh98,dh98,colangelo03,dty04,hocker04}.

The high accuracy achieved by the current experimental determination of
$a_\mu$~\cite{bnlgminus2} places a significant premium on 
reducing the error on $\left[ a_\mu\right]_{had}^{LO}$, 
which currently dominates the uncertainty on the Standard Model (SM)
prediction for $a_\mu$ (see Ref.~\cite{passera04} for a recent
review). At the desired
level of precision, the $\tau$ decay data can be used only
after taking into account the small 
isospin-breaking (IB) corrections to the CVC relation between
the charged and neutral current isovector spectral functions. A detailed
investigation of possible sources of such corrections,
for the numerically dominant $\pi\pi$ contribution, has
been made in Refs.~\cite{cen01,cen02}, and the resulting 
$s$-dependent IB correction factor incorporated into the latest $\tau$-based 
analyses~\cite{dehz03,hocker04} of $\left[ a_\mu\right]_{had}^{LO}$. 
A comparison of the corrected, $\tau$-based spectral data
with that obtained from the recent high-precision CMD-2 
experiment~\cite{cmd203}, however, shows significant residual disagreement
in the $\pi\pi$ components of the two versions of the isovector spectral
function~\cite{dehz03}:
the two are compatible below, and in the vicinity of, the
$\rho$ peak, but differ by $\sim 5-10\%$ for $m_{\pi\pi}$
between $\sim 0.85$ and $\sim 1$ GeV~\cite{dehz03}. This discrepancy leads to 
incompatible determinations of $\left[ a_\mu\right]_{had}^{LO}$,
the $\tau$-based determination lying $\sim 2\sigma$ 
higher~\cite{dehz03,hmnt04,hocker04}, and producing a SM prediction
for $a_\mu$ in agreement with the experimental result, while the
EM-based determination yields a SM prediction which
differs from experiment by $\sim 2.5\sigma$
(see Ref.~\cite{hocker04} and references therein for more details).

The preliminary KLOE $e^+e^-\rightarrow\pi^+\pi^-$ 
radiative return data~\cite{kloe04} supports the earlier EM-based
determination, yielding a value of $\left[ a_\mu\right]_{had}^{LO}$ 
compatible with that obtained using the CMD-2 $\pi\pi$ data~\cite{hocker04}.
However, as has been pointed out by many authors, the point-by-point 
agreement between the CMD-2 and KLOE cross-sections
is less than satisfactory~\cite{hocker04}, the KLOE data
lying higher than CMD-2 below the $\rho$ peak and lower than CMD-2 
both on the peak and above it. The structure of the weight $K(s)/s$ is such
that the effects of these discrepancies largely cancel
in $\left[ a_\mu\right]_{had}^{LO}$, but the situation nonetheless
remains unsatisfactory. 

Differences in the $\rho^0$ and $\rho^\pm$ masses and widths,
suggested as one possibility for resolving the $\pi\pi$ spectral
function discrepancy~\cite{gj03}, appear able to reduce locally, 
but not resolve fully the discrepancy~\cite{hocker04}.

Recent developments further complicate the picture. In Ref.~\cite{kmamu},
QCD sum rule constraints on the electroproduction and $\tau$ decay
data were investigated. Sum rules of the form
\begin{equation}
\int_{s_{th}}^{s_0}ds\, w(s)\, \rho(s)\, =\, 
{\frac{-1}{2\pi i}}\, \int_{\vert s\vert =s_0}ds\, w(s)\, \Pi (s)
\label{basicfesr}\end{equation}
were employed, where $\Pi (s)$ is either the EM or charged isovector
vector current correlator, $\rho (s)$ is the corresponding
spectral function, $s_{th}$ is the relevant threshold,
and $w(s)$ is a function analytic inside
and on the contour $\vert s\vert =s_0$. The OPE is employed
on the RHS, providing the desired constraints.
At the scales employed, the OPE for the vector current correlators
is essentially entirely dominated by the dimension $D=0$ perturbative
contribution, and hence determined by the single input, $\alpha_s$.
This input may be taken from high-scale determinations of $\alpha_s(M_Z)$
which are independent of the EM and $\tau$ data being tested. 
It turns out that both the normalization and $s_0$-dependence
of the weighted spectral integrals generated from
the hadronic $\tau$ decay data are in excellent agreement with
OPE expectations~\cite{kmamu}.  In contrast,
the weighted EM spectral integrals (obtained using CMD-2 data for
the $\pi\pi$ spectral component) do not agree with OPE
expectations, having (i) normalizations which are
$\sim 2\sigma$ low, and (ii) slopes with respect to $s_0$ 
which are $\sim 2.5\sigma$ low~\cite{kmamu}. These observations
suggest either a problem with the EM data, or the presence
of non-negligible non-one-photon physics contributions to the EM 
cross-sections. In either case, the results favor determinations of 
$\left[ a_\mu\right]_{had}^{LO}$ which incorporate hadronic 
$\tau$ decay data over those based
on EM data alone, and a SM prediction for $a_\mu$
in agreement with the current BNL experimental result~\cite{bnlgminus2}.
The recently-released SND $e^+e^-\rightarrow\pi^+\pi^-$ 
cross-section results~\cite{snd05pipi} are compatible with the 
IB-corrected $\tau$ data, and support this conclusion. 

In light of the above unsettled situation, we revisit the question 
of the reliability of the determination of the IB corrections 
which must be applied to the $\tau$ decay data,
focussing on two aspects of the existing treatment. 
We denote the correction to $\left[ a_\mu\right]_{had}^{LO}$ 
associated with these
IB corrections by $\left[\delta a_\mu\right]_{had}^{LO}$.

The first point concerns the uncertainty on the estimate for the
contribution to $\left[\delta a_\mu\right]_{had}^{LO}$ associated with 
``$\rho$-$\omega$ mixing''~\cite{footnote1} (present
in the EM, but not the $\tau$, spectral function).
The most recent updates of the $\tau$-based evaluation 
of $\left[ a_\mu\right]_{had}^{LO}$~\cite{dehz03,hocker04},
employ the IB corrections of Ref.~\cite{cen02} (CEN). The CEN
analysis is based on a version of the ChPT-constrained 
model for $F_\pi (s)$ developed by Guerrero and Pich~\cite{gp97} (GP).
The original GP model, which involved only the isospin conserving (IC) 
component of $F_\pi (s)$, was modified by CEN through the addition of
an IB $\omega\rightarrow\pi\pi$ contribution having the nominal 
$\rho$-$\omega$ mixing form. 
We refer to the resulting model as the GP/CEN model.
Using the parameter values given by CEN, that part of the full
IB correction associated with $\rho$-$\omega$ mixing becomes
\begin{equation}
\left[\delta a_\mu\right]_{had;mix}^{LO} 
= \left( 3.5\pm 0.8\right)
\times 10^{-10}\ ,
\label{cenrhoomega}\end{equation}
with the quoted uncertainty due essentially entirely to the
$20\%$ uncertainty on the parameter $\theta_{\rho\omega}$, which
describes the overall strength of $\rho$-$\omega$ ``mixing'' 
in the model. The uncertainty in Eq.~(\ref{cenrhoomega}) 
represents only a minor component of the total $\pm 2.6\times 10^{-10}$
uncertainty quoted by CEN for the full set of IB corrections~\cite{cen02}.

The GP/CEN model, however, is not the only one available for $F_\pi (s)$.
The Gounaris-Sakurai (GS) model~\cite{gs68}, the Kuhn-Santamaria (KS)
model~\cite{ks90}, and the hidden local symmetry (HLS) 
model~\cite{hlsbasic}, for example, all predate the GP/CEN model and
have been used extensively in the literature. The models differ
in the form employed for the broad IC component of the 
$e^+e^-\rightarrow\pi^+\pi^-$ amplitude, which is given (or dominated) by 
the $e^+e^-\rightarrow\rho^0\rightarrow\pi^+\pi^-$ contribution.
Implicit in the CEN error estimate is the (given the narrowness of
the $\omega$, plausible) assumption that the value obtained for
$\left[\delta a_\mu\right]_{had;mix}^{LO}$ will 
be largely insensitive to 
which of the models is employed in extracting the interference signal.
The high level of cancellation in the
$K(s)/s$-weighted integrals of the interference components of the various
model cross-sections, however, makes 
$\left[\delta a_\mu\right]_{had;mix}^{LO}$
{\it much} more model-dependent than would be naturally
anticipated. The resulting theoretical systematic error turns out to 
significantly exceed that associated with uncertainties on
the fitted model parameters for any given model, including
the GP/CEN model. This point is discussed in more detail in section
\ref{sec:rhoomegaerror} below.

\begin{figure}
\epsfig{figure=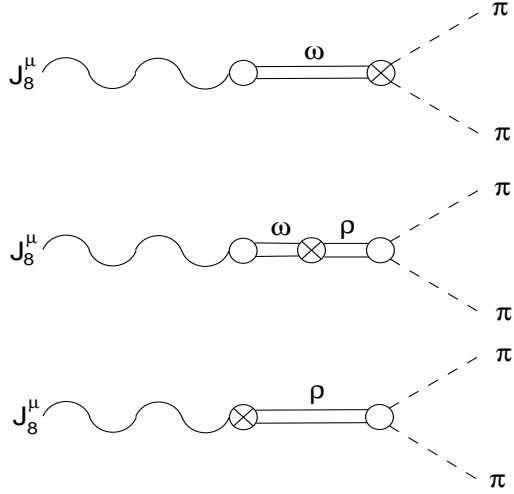, height=12.5cm,width=11cm}
\caption{\label{fig1}Isospin-breaking resonance contributions
to $e^+e^-\rightarrow\pi^+\pi^-$}
\end{figure}

The second point concerns an IB correction not accounted for in the 
CEN analysis. In the limit that (as for the IC component) the
IB component of the $e^+e^-\rightarrow\pi^+\pi^-$ amplitude is assumed 
dominated, away from threshold, by resonance contributions, three such
contributions will, in principle, be present in the
$\rho$, $\omega$ resonance region. These are shown in Fig.~\ref{fig1}, 
where the open circles represent IC vertices
and the crossed circles IB vertices. $J_\mu^3$ and $J_\mu^8$
are the isovector and isoscalar members of the vector current
octet. The
first two graphs represent the $\rho$-$\omega$ mixing and direct IB 
$\omega\rightarrow\pi\pi$ decay contributions to the amplitude.
They are small away from the $\omega$ peak region, generate
contributions to the flavor `38' part of the 
EM spectral function, and combine to produce the prominent narrow 
interference shoulder in the experimental cross-section. 
The remaining graph depicts the contribution associated with
the IB (isoscalar) component of the $\rho^0$ EM decay constant.
Such a component of the decay constant 
is unavoidable in the SM.

Because of the narrowness of the $\rho$-$\omega$ interference shoulder, 
the interference part of the cross-section can, modulo the model dependence 
noted above, be determined experimentally. The corresponding contribution to
$\left[\delta a_\mu\right]_{had}^{LO}$ can thus also, with the same caveat, 
be determined experimentally. In contrast, the interference
contribution associated with the isoscalar $\rho^0$ EM decay constant
(which also belongs to the flavor `38' part of the cross-section)
is identical in shape to the dominant, broad, IC flavor `33' 
contribution and not (even in principle)
extractable experimentally. The corresponding contribution to 
$\left[\delta a_\mu\right]_{had}^{LO}$, which is certainly present 
at some level in the SM, has not, to our knowledge, been investigated
in the literature, and certainly is {\it not} included
in the treatment of IB corrections employed by CEN, for the
reason explained below. It is, in fact,
analogous to the $\rho$-$\omega$ ``mixing'' contribution, which was
also not present in the GP model approach~\cite{cen01}, and
hence had to be added by hand to the GP model expression
by CEN~\cite{cen02}.

The reason the three IB resonance contributions to $F_\pi (s)$
shown in Fig.~\ref{fig1} are not
incorporated in the GP model framework is as follows. 
The GP model is constructed by implementing the constraints of
unitarity, analyticity and short-distance QCD,
and requiring that the model expression for $F_\pi (s)$ 
match properly onto the known next-to-leading 
order (NLO) ChPT expression at low energy. This last
constraint is realized using the resonance chiral effective 
theory approach, in which low-energy resonance effects appear
through contributions proportional
to the NLO low energy constants (LEC's), $L_k^r$, 
of Gasser and Leutwyler~\cite{gl85}. It is, however, straightforward
to demonstrate that, at NLO, $\langle \pi^+\pi^- \vert J_\mu^8\vert 0\rangle$
receives contributions only from loops, and not from the NLO 
LEC's~\cite{omtw97}. As a result, none of the resonance-induced IB effects 
depicted in the figure are incorporated in the GP expression
for $F_\pi (s)$. Although, numerically, the IB loop effects are tiny 
near threshold~\cite{omtw97}, the obvious experimental 
interference shoulder in the $\rho$-$\omega$ region
shows that this does not remain the case at higher energies.
In order to include $\rho$-$\omega$ interference, it was thus
necessary for CEN to
add a $\rho$-$\omega$ ``mixing'' contribution to the GP model
by hand. The broad IB $\rho$ contribution is similarly absent 
in the GP model approach, and would also have to be added by hand.
In Section~\ref{sec:ibfpi} we investigate sum rule constraints on this
contribution, and, in addition, use the size of the analogous effect
in the pseudoscalar sector, as evaluated at NLO
in the chiral expansion, to obtain some guidance as to what 
the natural scale of the effect might be.

A brief summary, and our conclusions, is given in 
Section~\ref{sec:conclusions}.

\section{\label{sec:rhoomegaerror}Model Dependence of 
$\left[\delta a_\mu\right]_{had;mix}^{LO}$}

The pion form factor, in the GS model, is given by~\cite{gs68}
\begin{equation}
F_{\pi}^{\rm(GS)}(s) = {\frac{1}{(1+\beta )}}
       \left({\rm BW}_\rho^{(GS)}(s)
\left[1 + \delta{\frac{s}{m_{\omega}^2}}
P_\omega (s)\right] + 
  \beta {\rm BW}_{\rho^\prime}^{(GS)}(s)\right )
\end{equation}
where
\begin{eqnarray}
P_\omega (s) & = & {\frac{m_{\omega}^2}
     {(m_\omega^2-s-{\rm i}m_{\omega}\Gamma_{\omega})}} \nonumber\\
{\rm BW}_V^{(GS)}(s) & = & 
         {\frac {m_V^2\left(1+d(m_V){\frac{\Gamma_V}{ m_V}}\right)}
{\left(m_V^2-s+f(s,m_V,\Gamma_V)-{\rm i}m_V\Gamma_V(s,m_V,\Gamma_V)
\right)}}\end{eqnarray}
with
\begin{eqnarray}
d(m_V) & = & {\frac{3m_{\pi}^2}{(\pi p_{\pi}^2(m_V^2))}}\, \ell n\left(
{\frac{(m_V+2p_{\pi}(m_V^2))}{2m_{\pi}}}\right) + 
{\frac{m_V}{(2\pi p_{\pi}(m_V^2))}} -
{\frac{m_{\pi}^2 m_V}{(\pi p_{\pi}^3(m_V^2))}} \nonumber\\
f(s,m_V,\Gamma_V) & = & {\frac{\Gamma_V m_V^2}{p_{\pi}^3(m_V^2)}}\left(
   p_{\pi}^2(s)[H(s)-H(m_V^2)] + 
   (m_V^2-s)p_{\pi}^2(m_V^2){\frac{dH}{ds}}(m_V^2)\right) \nonumber\\
H(s) & = & {\frac{2 p_{\pi}(s)}{\pi\sqrt{s}}}\, \ell n\left(
     {\frac{\sqrt{s} + 2 p_{\pi}(s)}{2 m_{\pi}}}\right ) 
\end{eqnarray}
where $p_{\pi}(s) = \sqrt{{{s}\over {4}}-m_{\pi}^2}$
is the pion CM momentum for squared invariant mass $s$,
$\Gamma_V(s,m_V,\Gamma_V)$ is the standard $s$-dependent width
for vector meson $V$ implied by $p$-wave phase space, and
$\Gamma_V = \Gamma_V(m_V^2,m_V,\Gamma_V)$.

Similarly, for the KS model, one has~\cite{ks90}
\begin{equation}
F_{\pi}^{\rm (KS)}(s) = \left( {\frac{P_{\rho}(s)\left (
{\frac{1 + \delta P_{\omega}(s)}{1+\delta}}\right) 
+ \beta P_{\rho^{\prime}}(s) + 
\gamma P_\rho^{\prime\prime}(s)}{1+\beta+\gamma}} \right )
\end{equation}
with 
\begin{equation}
P_V(s)  =  {\frac{m_V^2}{m_V^2-s-{\rm i}m_V\Gamma_V(s,m_V,\Gamma_V)}}\ .
\end{equation}

The HLS model~\cite{hlsbasic}, as implemented by CMD-2, has the form
\begin{equation}
F_{\pi}^{(HLS)}(s) = 1-{\frac{a_{HLS}}{2}}+{\frac{a_{HLS}}{2}}\left({\frac
{P_{\rho}(s) \left( 1 + \delta P_{\omega}(s)\right)}
{1+\delta}} \right )
\end{equation}
with $a_{HLS}$ a constant. The model provides a good quality fit to the
data below $1$ GeV despite having no explicit
$\rho^\prime$ contribution. It also turns out to reproduce
the correct final state $\pi\pi$ phases after the model
parameters have been fitted~\cite{hlsgood}.

For all of the GS, KS and HLS models, the constant $\delta$, which
parametrizes the strength of the narrow IB 
amplitude, is taken to be complex. A non-zero phase 
is, in general, unavoidable in the presence of an
IB direct $\omega\rightarrow\pi\pi$ decay contribution~\cite{footnote1}.

The GP model for the pion form factor is given by~\cite{gp97}
\begin{equation}
F^{(GP)}_\pi (s) = 
P_\rho (s) \,  \exp\left(
{\frac{-s}{96 \pi^2 f_\pi^2}}
\left[ {\rm Re}\, 
L\left({\frac{m_\pi^2}{s}},{\frac{m_\pi^2}{m_\rho^2}}\right)\, +\, 
{\frac{1}{2}}\, {\rm Re}\, L\left( {\frac{m_K^2}{s}},
{\frac{m_K^2}{m_\rho^2}}\right)
\right]\right)\ ,
\label{gpmodel}\end{equation}
where
\begin{equation}
L\left( {\frac{m^2}{s}},{\frac{m^2}{m_\rho^2}}\right) = 
\ell n\left( {\frac{m^2}{m_\rho^2}}\right) + {\frac{8m^2}{s}} -{\frac{5}{3}}
+ \beta (s)^3 \, \ell n\left[ {\frac{\beta (s)+1}{\beta (s)-1}}\right]
\end{equation}
with $\beta (s) = \sqrt{ 1- 4m^2/s}$ and the $s$-dependent width,
$\Gamma_\rho (s,m_\rho ,\Gamma_\rho )$ appearing in $P_\rho (s)$ 
replaced by the resonance chiral effective theory expression
\begin{equation}
\Gamma_\rho (s) = {\frac{m_\rho s}{96 \pi f_\pi^2}}\left(
\theta (s-4m_\pi^2)\, \beta_\pi (s)^3 \, +\,
{\frac{1}{2}} \theta (s-4m_K^2)\, \beta_K(s)^3\right)\ .
\label{gpwidth}\end{equation}
Some IB effects are incorporated
into $F^{(GP)}_\pi (s)$ if one evaluates the
phase space factors in the $s$-dependent width 
using the physical charged $\pi$ and $K$ masses.
In Ref.~\cite{cen02}, a small rescaling of the coefficient appearing on 
the RHS of Eq.~(\ref{gpwidth}) is allowed in order to account for
the $\sim 1.5$ MeV contribution of $\pi\pi\gamma$ decays
to the total width of the $\rho$~\cite{cen02,vcnote}.
The CEN modification of $F^{(GP)}_\pi (s)$, designed to
incorporate the $\rho$-$\omega$ mixing contribution not
included in the original GP model, then has the form
\begin{equation}
F^{(GP/CEN)}_\pi (s) = F^{(GP)}_\pi (s) - P_\rho (s)
\left({\frac{\theta_{\rho \omega}}{3m_\rho^2}}\right)
\left({\frac{s}{m_\omega^2}}\right) P_\omega (s)\ .
\label{gpcenmodel}\end{equation}
The parameter $\theta_{\rho \omega}$ was assumed real by
CEN. 

The original version of the GP/CEN model, as parametrized by CEN, predates 
the most recent, corrected version of the CMD2 data, 
and does not provide a good fit to it, producing a $\chi^2$ of $80$ for the
$43$ CMD-2 data points. The fit quality can be improved 
by allowing $m_\rho$, $\theta_{\rho \omega}$ (still assumed
real) and the rescaling of the resonance chiral
effective theory width to be fit to data, but the
resulting optimized fit still has a $\chi^2$
of $61$ for the resulting $40$ degrees of freedom. 
The corresponding mixing contribution, 
$\left[\delta a_\mu\right]_{had;mix}^{LO}$,
is shifted only slightly from the original CEN value,
from $3.5\times 10^{-10}$ to $3.7\times 10^{-10}$.
The fact that the $\rho$-$\omega$ ``mixing'' signal
is actually a combination of mixing and direct $\omega\rightarrow\pi\pi$
effects, however, means that an effective representation for the combination 
of the form given by Eq.~(\ref{gpcenmodel}) is not generally
possible without allowing 
$\theta_{\rho \omega}$ to have a non-zero phase~\cite{mow}. 
If we extend the GP/CEN model in this way, treating the phase as a
fourth parameter to be fit to the data,
an acceptable fit, having $\chi^2 =41$ 
for the remaining $39$ degrees of freedom, becomes possible. 
We refer to this version of the GP/CEN model as GP/CEN$^*$
in the table below.

The values obtained for 
$\left[\delta a_\mu\right]_{had;mix}^{LO}$ 
in the various models are given in Table~\ref{table1}. All
results are generated using versions of the models optimized
to the most recent versions of the CMD-2~\cite{cmd203} {\it bare} 
cross-section data~\cite{footnote2}. Only the GP/CEN$^*$
version of the GP/CEN model is included since the unmodified
version does not produce an acceptable quality fit.

\begin{table}
\caption{\label{table1}
$\left[\delta a_\mu\right]_{had;mix}^{LO}$
for the various models discussed in the text fit to 
the most recent CMD-2 bare $e^+e^-\rightarrow \pi\pi$ 
cross-sections}
\vskip .1in
\begin{tabular}{lcc}
\hline
Model&\qquad\qquad $\chi^2$/dof\qquad\qquad&
$\left[\delta a_\mu\right]_{had;mix}^{LO}
\times 10^{10}$\\
\hline
GS&\ \ \ \ 36/38&$2.0\pm 0.5$\\
HLS&\ \ \ \ 37/38&$4.0\pm 0.6$\\
KS&\ \ \ \ 37/38&$3.9\pm 0.6$\\
GP/CEN$^*$&\ \ \ \ 41/39&$2.0\pm 0.5$\\
\hline
\end{tabular}
\end{table}

Two things are evident from the table. First, the sensitivity
of $\left[\delta a_\mu\right]_{had;mix}^{LO}$ to the model
employed is much larger than that associated with the
uncertainties on the values of the fitted model parameters for a 
given model.
Second, by comparing the GP/CEN$^*$ results to those for the GP/CEN 
version of the model (quoted above), we see that allowing a non-zero phase 
for IB parameter $\theta_{\rho\omega}$ leads to a significant decrease
in $\left[\delta a_\mu\right]_{had;mix}^{LO}$. (A similar effect is
produced by the phase of $\delta$ in the other models.)
The origin of these effects is easy to understand. In the interference
region, the IC amplitude is $\simeq B_\rho (s)$, where $B_\rho (s)$
is the $\rho (770)$ Breit-Wigner-like form in the given model. Writing
the IB amplitude in the ``$\rho$-$\omega$ mixing''
form, generically $B_\rho (s) \delta P_\omega$, with 
$\delta =\vert\delta\vert e^{i\phi}$,
the flavor '38' component of the EM cross-section is
then given approximately by the expression
\begin{equation}
\vert B_\rho (s)\vert^2\left[{\frac{2\, \vert\delta\vert\, m_\omega^2}
{\left[ \left( m_\omega^2-s\right)^2+m_\omega^2\Gamma_\omega^2\right]}}
\, \left( cos(\phi )\, (m_\omega^2-s)\, -\, m_\omega\Gamma_\omega
sin(\phi )\right)\right]\ .
\label{ibicform}\end{equation}
Since the coefficient multiplying $cos(\phi )$ in the square bracket of
Eq.~(\ref{ibicform}) is antisymmetric about $s=m_\omega^2$, 
the corresponding contribution to 
$\left[\delta a_\mu\right]_{had;mix}^{LO}$ 
vanishes in the limit that one neglects the variation of 
$\vert B_\rho (s)\vert^2\, K(s)/s$ over the $\omega$ region.
Since both $\vert B_\rho (s)\vert^2$ and $K(s)/s$ are
decreasing functions in this region, a small residual positive
contribution remains. The coefficient of $sin(\phi )$, in contrast,
is symmetric, so no analogous cancellation is present in the corresponding
contribution to $\left[\delta a_\mu\right]_{had;mix}^{LO}$,
i.e., the $sin(\phi )$ integral is strongly enhanced relative to
the $cos(\phi )$ integral. It is the strong cancellation in the 
$cos(\phi )$ integral, 
combined with small differences in the $s$-dependence of 
$B_\rho (s)$ in the different models, which accounts for the significant 
model dependence in the results 
for $\left[\delta a_\mu\right]_{had;mix}^{LO}$.
Since fits to the data favor small positive $\phi$ for all
the models considered here, the $sin(\phi )$ contribution to
$\left[\delta a_\mu\right]_{had;mix}^{LO}$ is negative.
The relative enhancement of the $sin(\phi )$ integral
means that the cancellation against the $cos(\phi )$ contribution, 
which is absent if one sets $\phi$ to zero from the outset, 
can be quite significant, even for relatively small $\phi$.

Table~\ref{table2} shows the impact of the choice of input data set
on $\left[\delta a_\mu\right]_{had;mix}^{LO}$, giving
the central values corresponding to optimized fits of each model 
to the bare CMD-2 and SND cross-sections. In the case of the KLOE
data, the optimized fits for all four models have 
$\chi^2/dof >2$~\cite{footnote4}; the corresponding
$\left[\delta a_\mu\right]_{had;mix}^{LO}$ results have
therefore been omitted from the table. A $\chi^2/dof >2$ is
also obtained for the optimized fit of the GP/CEN$^*$ model
to the SND data; the corresponding entry in the table,
though included for completeness, has been
enclosed in parentheses to remind the reader of this fact~\cite{footnote3}. 
From the table we see that the variation 
in $\left[\delta a_\mu\right]_{had;mix}^{LO}$ values
among the different models, for a fixed input data set,
is significantly larger than the variation of the results for a given
model over the different input data sets. It is
thus the theoretical systematic error associated with 
choice of model used in separating the IC and IB components of
the amplitude which dominates the uncertainty in the
determination of $\left[\delta a_\mu\right]_{had;mix}^{LO}$.

\begin{table}
\caption{\label{table2}Central values of 
$\left[\delta a_\mu\right]_{had;mix}^{LO}$,
in units of $10^{-10}$, for the optimized fits of the
various models to the {\it bare} CMD-2~\protect\cite{cmd203}
and SND~\protect\cite{snd05pipi} $e^+e^-\rightarrow\pi\pi$ cross-sections. } 
\vskip .1in
\begin{tabular}{lcc}
\hline
Model&\qquad CMD-2\qquad&\qquad SND\qquad\\
\hline
GS&\ \ 2.0&\ \ \ 2.2\\
HLS&\ \ 4.0&\ \ \ 4.5\\
KS&\ \ 3.9&\ \ \ 4.3\\
GP/CEN$^*$&\ \ 2.0&\ \ \ (1.6)\\
\hline
\end{tabular}
\end{table}

For comparison, and to further illustrate the sensitivity of the mixing
contribution to small changes in the data and the resulting
model fits, results corresponding to some older fits from the literature
are given in Table~\ref{table3}. `CEN02' labels the original GP/CEN 
result~\cite{cen02} (with no phase for the parameter $\theta_{\rho\omega}$), 
`ALEPH97' the results corresponding to the ``combined'' ($\tau$ plus
electroproduction) GS and KS fits of Ref.~\cite{aleph97} 
and `DAVIER03' the results corresponding to the 
similarly ``combined'' GS fit (Table 4) of Ref.~\cite{davier03}. 
Details of the fit procedures, and data sets employed 
may be found in the original references. The total $\chi^2$ of
the fits {\it relative to the 2003 CMD-2 data} are also given.
The quoted results correspond in each case to the central values of
the fit parameters for the models.
Typically, differences between the old and new values of
$\left[\delta a_\mu\right]_{had;mix}^{LO}$ for a given model
are much larger than one might have anticipated, given the relatively
small changes in the both data and fitted parameter values.
This sensitivity is again a reflection of the 
strong cancellation in the integral of the product of IC and
IB amplitudes.

\begin{table}
\caption{\label{table3}Central values for
$\left[\delta a_\mu\right]_{had;mix}^{LO}$
for earlier model fits in the literature}
\begin{tabular}{llcc}
\hline
Model&Reference&\qquad $\chi^2$/dof\qquad &\qquad
$\left[\delta a_\mu\right]_{had;mix}^{LO}
\times 10^{10}$\qquad\\
\hline
GS&ALEPH97&\ \ \ \ 59/38&$4.3$\\
KS&ALEPH97&\ \ \ \ 75/38&$6.3$\\
GS&DAVIER03&\ \ \ \ 65/38&$2.5$\\
GP/CEN&CEN02&\ \ \ \ 80/42&$3.5$\\
\hline
\end{tabular}
\end{table}

We conclude that the inability to separate the IC and IB components of the 
$e^+e^-\rightarrow\pi\pi$ amplitudes in a model independent manner
leads to a significant uncertainty in the 
evaluation of $\left[\delta a_\mu\right]_{had;mix}^{LO}$.
How one assesses this uncertainty depends on one's attitude to the
various models. One stance might be to argue that the GS and GP/CEN models,
which explicitly incorporate the constraints of unitarity and
analyticity, are to be favored in deciding on a central value. The variation
of the results across the different models would then
serve as a measure of the residual uncertainty. Alternately, since
all of the models are, to greater or lesser extent, phenomenological,
and, at least for the CMD-2 data, yield comparable quality fits,
one could instead average the results to arrive
at a central value, and assign an error large enough to incorporate
the highest and lowest values allowed by the errors associated
with those on the fitted parameters for
the various models. The first stance would yield
\begin{equation}
\left[\delta a_\mu\right]_{had;mix}^{LO} \, =\, 
(2.0\pm 2.9)\times 10^{-10}\ ,
\label{stance1}\end{equation}
the second
\begin{equation}
\left[\delta a_\mu\right]_{had;mix}^{LO} \, =\, 
(3.1\pm 1.8)\times 10^{-10}\ .
\label{stance2}\end{equation}
A smaller central value, and significantly larger uncertainty, would result
if one ignored the poor quality of the model
fits and also took the KLOE-based results into account~\cite{footnote5}.

\section{\label{sec:ibfpi}The Broad IB $\rho$ Contribution}

Although the broad IB $\rho$ contribution to the experimental
$e^+e^-\rightarrow\pi^+\pi^-$ cross-section has the same shape
as the dominant IC contribution,
and hence cannot be separated
from it experimentally, it can,
in principle, be determined theoretically through a QCD sum
rule analysis of the IB vector current correlator, $\Pi^{38}(q^2)$,
defined by
\begin{equation}
i\, \int\, d^4x\, e^{iq\cdot x}\,
\langle 0\vert T\left( J_\mu^3(x) J_\nu^8(0)\right)\vert 0\rangle
\equiv \left( q_\mu q_\nu -q^2 g_{\mu\nu}\right)\, \Pi^{38}(q^2)\ .
\label{ibcorrdef}\end{equation}
A non-zero (IB) coupling
of the $\rho$ to $J_\mu^8$ will produce a broad $\rho$ contribution to the
spectral function of $\Pi^{38}$ , $\rho^{38}(s)$, 
whose strength, $X_\rho$, is proportional to the
product of the IC isovector and IB isoscalar $\rho^0$ decay constants.
This product, together with analogous IB products, $X_V$, for the other
vector meson resonances, can, in principle, be determined by matching the
appropriate weighted integrals of $\rho^{38}$ 
to the corresponding OPE expressions.

In Ref.~\cite{mw98} such an analysis was performed using two different
families of ``pinch-weighted'' finite energy sum rules (pFESR's)~\cite{kmsr98}.
The details of the analysis may be found in Ref.~\cite{mw98}, and
will not be repeated here. The following point is, however, worth
noting. Because the numerically dominant term on the OPE side
of the various sum rules is that with dimension $D=4$, the contribution
from the VEV's of $D=6$ four-quark operators, $c_6O_6$, is not expected to be
negligible. The fact that the VEV's for such operators are typically
not well known empirically would normally present
a problem for the sum rule analysis. 
It turns out that the dependence of the $X_V$ 
on $c_6O_6$ is different for the two different pFESR families,
allowing, not only the $X_V$, but also $c_6O_6$, to be determined
from the combined analysis. The values of $c_6O_6$ which make the 
different $X_V$ consistent turn out to agree at the $\sim 1\%$ level,
providing strong support for the reliability of the analysis. 

Unfortunately, in the analysis of Ref.~\cite{mw98}, the 
$\rho$-$\omega$ ``mixing'' contribution to $\rho^{38}(s)$ implied by 
the observed interference shoulder in the EM cross-section
was not input separately on the spectral side of the sum rules employed.
As a result, the output $X_\rho$ contains contributions from all three of the 
IB processes shown in Figure~\ref{fig1}. In order to separate out
the experimentally inaccessible contribution associated with
the third of those processes, we
have redone the analysis of Ref.~\cite{mw98}, this time inputting
the ``interference'' component of $\rho^{38}$, as determined from the
CMD-2 experimental data in the interference region. This input,
as noted above, depends to some extent on
the choice of model for the $\rho$ contribution
to the IC component of the amplitude. We have then used the
various pFESR's to solve for the residual broad $\rho$ contribution,
which yields directly the contribution
to the flavor `38' part of the EM spectral function
associated with the third graph in Fig.~\ref{fig1}.
With current experimental errors, an accurate
determination of this ``direct'' contribution, and hence of the associated
contribution to $\left[\delta a_\mu\right]_{had}^{LO}$, 
$\left[\delta a_\mu\right]_{had;direct}^{LO}$,
turns out to be impossible. We find central values of a natural
scale (see below), but with errors, induced by the uncertainy
in the integrated (model-dependent) $\rho$-$\omega$ 
``interference'' term, much 
larger than these central values.
Even more unfortunately, versions of the analysis using 
errors scaled down artificially by hand suggest that
the reduction needed to allow even just a reliable determination
of the sign of the direct effect are unlikely to be reachable
in the foreseeable future.

In view of the weakness of the constraints arising from the
sum rule analysis, we turn to the pseudoscalar sector,
and study the size of analogous effects in the IB decay
constant, $f_\pi^8$, of the $\pi^0$, defined by
\begin{equation}
\langle 0\vert A_\mu^8 \vert \pi^0\rangle \, =\, i\, f_\pi\epsilon_1 q_\mu
\equiv i\, f_\pi^8q_\mu\ .
\label{ibfpi0}\end{equation}
At leading order in the chiral expansion, the IB parameter 
$\epsilon_1 = f_\pi^8/f_\pi$ is equal to 
$\theta_0=\sqrt{3}(m_d-m_u)/4(m_s-\hat{m})$, where $\hat{m}=(m_d+m_u)/2$,
and is due entirely to IB mixing on the external leg, induced by
the non-zero value of $m_d-m_u$. At NLO, $\epsilon_1$ 
receives contributions both 
from mixing and from IB in the low energy representation of the
axial current $A_\mu^8$. The full NLO expression for $\epsilon_1$
is given in Ref.~\cite{gl85}, while the ingredients necessary for
separating the mixing and vertex contributions may be found in
Ref.~\cite{kmpietamixing}. At NLO the separate mixing and vertex
contributions are, as expected on physical grounds, scale dependent.
The (scale-dependent) ``direct'' (vertex)
contribution is given by
\begin{eqnarray}
\left[{\frac{f_\pi^8}{f_\pi}}\right]_{direct} \,& =&\,
\left( {\frac{16m_K^2 + 4m_\pi^2}{f_\pi^2}}\right) L_4^r(\mu ) 
+8{\frac{m_\pi^2}{f_\pi^2}} L_5^r(\mu )	- 
{\frac{m_K^2\, \ell n(m_K^2/\mu^2)}{8 \pi^2 f_\pi^2}}\nonumber\\
&&\qquad 
+\left({\frac{m_K^2-m_\pi^2}{12 \pi^2 f_\pi^2}}\right) [1+\ell n(m_K^2/\mu^2)]
\ ,
\label{psanalogue}\end{eqnarray}
where $\mu$ is the ChPT renormalization scale, and $L_{4,5}^r(\mu )$
are the renormalized NLO LEC's, evaluated at scale $\mu$~\cite{gl85}.

Taking, for illustration, the central values for $L_{4,5}^r$ from the main fit
of Ref.~\cite{abt01}, one finds that the leading (NLO) contribution
to $f_\pi^8$ varies from $0.58\%$ to $0.46\%$ of $f_\pi$ as
$\mu$ varies from $m_\eta$ to $1$ GeV. Using the slightly modified
fit denoted ``fit D'' in Ref.~\cite{bijnens04}, which produces 
good values for the threshold parameters for $\pi\pi$ and $\pi K$ 
scattering, the ratio runs from $0.60\%$ to $0.71\%$ over the
same range of $\mu$. We thus conclude that the NLO ``direct'' contribution
to the IB decay constant ratio is $\sim 1/2\%$ for typical
hadronic scale choices. A similar value for the corresponding
IB ratio of $\rho^0$ decay constants would produce a contribution
$\left[\delta a_\mu\right]_{had;direct}^{LO}\simeq (2-3)\times 10^{-10}$.
The central value obtained from the sum rule analysis turns
out to be comparable to, or less than, this natural size for all four models
considered here. The upper bound implied by the errors, however,
is much larger, preventing us from using the sum rule
constraints in a meaningful way.

Note that the presence of 
a small IB component in the $\rho^0$ EM decay constant
would have an impact on the values of $m_\rho$ and $\Gamma_\rho$
obtained by fitting the various phenomenological models to data.
At the natural scale (given above) for this effect, however, the
(model-dependent) 
shift in $m_\rho$ would be $\sim 0.2$ MeV or less, and that in
$\Gamma_\rho$ $0.4$ MeV or less. The effect, while contributing
to the apparent difference in charged and neutral $\rho$ masses
and widths, can thus account for at best a modest fraction
of the differences obtained after fitting the models independently
to the $\tau$ decay and electroproduction data.

\section{\label{sec:conclusions}Conclusions}

We have shown that the model dependence encountered
in separating the IB from the IC component of the 
$e^+e^-\rightarrow\pi^+\pi^-$ cross-section leads to a theoretical
systematic uncertainty on the $\rho$-$\omega$ ``mixing''
contribution to $\left[ \delta a_\mu\right]_{had}^{LO}$
which is several times that associated with
the fit parameter uncertainties for any given model.
There is also a potentially non-negligible IB contribution,
associated with the direct IB coupling of the isoscalar
part of the EM current to the $\rho^0$, which cannot,
even in principle, be determined experimentally.
At present, we are able only to make a rough ``natural size''
estimate for the magnitude of this contribution. Both of these effects
would need to be taken into account when estimating the
uncertainty in the IB correction required in order to incorporate 
the $\tau^-\rightarrow\nu_\tau \pi^-\pi^0$ spectral data
into the evaluation of $\left[ a_\mu\right]_{had}^{LO}$. 
The results make it clear that evaluating the ``mixing'' component
of the IB correction using only a single model in the analysis
of the cross-section will lead to a significant underestimate 
in the uncertainties.

In view of the results of the tests involving independent high-scale OPE
constraints reported in Ref.~\cite{kmamu}, and the compatibility of the
SND and $\tau$ $\pi\pi$ data sets, we base our final results
on the SND data set. (The CMD-2 and KLOE values,
quoted above, allow alternate choices to be made.) Choosing
the GS model as the favored case would lead to a decrease of
$1.3\times 10^{-10}$ in $\left[ \delta a_\mu\right]_{had;mix}^{LO}$,
relative to the standard CEN value. Such a shift would 
{\it lower} the $\tau$-based prediction for 
$\left[ \delta a_\mu\right]_{had}^{LO}$ in the SM, slightly
increasing the difference between the SM
prediction and the central experimental value, but leaving
them compatible at the $1\sigma$ level. For a given model,
the results for $\left[ \delta a_\mu\right]_{had;mix}^{LO}$
obtained using either the CMD-2 or SND data are actually in good
agreement. The decrease in central value relative to CEN would
thus also reduce the discrepancy between the CMD-2 and $\tau$-based
determinations of the the $\pi\pi$ contribution to 
$\left[ \delta a_\mu\right]_{had}^{LO}$ by $\sim 10\%$.
The HLS model, which produces an optimized fit to the SND data of comparable
quality to that of the GS model, in contrast, shifts the 
mixing contribution {\it up} by $1.0\times 10^{-10}$, relative to
the CEN value. Since the GS model has the
constraints of analyticity and unitarity explictly built into it,
while the HLS model does not, we have favored the GS model in our
discussion of the central value above. It is, however, important
to bear in mind that, in terms of quality of fit to the SND
data, the models cannot be distinguished, and hence that a significant
uncertainty, associated with the model dependence, 
must be attached to any particular chosen central value.
The range of values allowed by the CMD-2 data is fully contained
within that allowed by the SND data, but the latter range would
need to be significantly extended, including to negative
values of $\left[ \delta a_\mu\right]_{had;mix}^{LO}$, if the
lower-quality fits to the KLOE data were also taken into account.

Finally, taking into account the uncertainties already identified by
CEN, and adding to these both the increased uncertainty on
$\left[ \delta a_\mu\right]_{had;mix}^{LO}$ and a possible direct
IB isoscalar $\rho^0$ EM coupling contribution of the
``natural size'' discussed above, we arrive at a combined 
uncertainty for the IB correction one must apply in order to use the
$\tau^-\rightarrow\nu_\tau\pi^-\pi^0$ data in computing 
$\left[ \delta a_\mu\right]_{had}^{LO}$ which is $\sim 4\times 10^{-10}$.
It appears unlikely that this uncertainty can
be significantly reduced. Should the new BNL experimental
proposal~\cite{bnlproposal} be approved, 
the uncertainty on the IB corrections would thus exceed those 
on the experimental determination of $a_\mu$, seriously limiting
our ability to make use of hadronic $\tau$ decay data in
determining the SM prediction for
$\left[ \delta a_\mu\right]_{had}^{LO}$.

\begin{acknowledgments}
KM would like to acknowledge the hospitality of the CSSM, University of
Adelaide, and the ongoing support of the Natural Sciences and Engineering 
Council of Canada.
\end{acknowledgments}



\begin{thebibliography}{99}
\bibitem{footnote1}The 
IB $e^+e^-\rightarrow\omega^0\rightarrow\pi^+\pi^-$ amplitude,
whose interference with the leading isospin-conserving component, 
dominated by intermediate $\rho^0$ exchange, 
is responsible for the observed interference shoulder
in the experimental cross-section, is in fact generated
by both $\rho$-$\omega$ mixing and direct IB $\omega\rightarrow\pi\pi$
coupling effects. For a given model of the $\rho$
propagator, the experimentally-extracted Orsay phase can be used, in principal,
to separate these two contributions~\protect\cite{mow}. The 
recent CMD-2 data~\protect\cite{cmd203} shows clear evidence for
a non-zero value of the direct IB $\omega\rightarrow\pi\pi$ coupling,
as does the recent SND data~\protect\cite{snd05pipi}.
At leading order in IB, the two effects both generate 
contributions to the flavor '38' part of the
EM spectral function.
\bibitem{footnote2}Davier~\protect\cite{daviernote}, 
working with the GS model, 
has noted that the $m_\rho$ and $\Gamma_\rho$ values
obtained from fits to the bare and vacuum-polarization-dressed versions 
of the CMD-2 cross-sections differ by $\sim 1$ MeV. Similar shifts
also occur for the other models. These shifts produce corresponding
shifts in $\left[\delta a_\mu\right]_{had;mix}^{LO}$.
The versions associated with the fits to the bare cross-sections,
quoted in the text, lie higher than those
associated with the fit to the vacuum-polarization-dressed values 
by $(0.4\rightarrow 0.6)\times 10^{-10}$, depending on the model.
The relevant versions for comparison to the $\tau$ data are,
as explained by Davier, those obtained from the
fit to the bare cross-sections~\protect\cite{daviernote}.
\bibitem{footnote4}The fits to the KLOE data all require 
significantly larger values of the phase of $\delta$ (or
$\theta_{\rho\omega}$) than do the corresponding fits to the
CMD-2 and SND data sets. Such phase values lead to significantly lower
$\left[\delta a_\mu\right]_{had;mix}^{LO}$,
for the reasons explained in the text. In view of the poor quality of 
the corresponding fits, we have, however, excluded these results from
consideration in arriving at a final assessment of the range of allowed values 
for $\left[\delta a_\mu\right]_{had;mix}^{LO}$.
Were one to include them, the range would be considerably expanded,
extending even to small negative values.
\bibitem{footnote3}Adding an explicit $\rho^\prime$ contribution to the
GP/CEN$^*$ model would, of course, allow one to obtain
an improved fit to the SND data. Since the original CEN version
did not include such a contribution, we do not tabulate results
for such a further-modified version of the GP model here.
\bibitem{footnote5}The optimized fits to the KLOE data yield
$\left[\delta a_\mu\right]_{had;mix}^{LO}\,\times 10^{10} =\, 
-0.2\pm 0.7,\ -0.6\pm 0.8,\ 0.6\pm 0.8,$ and $-0.6\pm 0.7$ for
the GS, KS, HLS and GP/CEN$^*$ models, respectively.
\end{thebibliography}
\end{document}